\documentclass[aip,rsi,amsmath,amssymb,
reprint,longbibliography%
]{revtex4-1}

\usepackage{graphicx}
\usepackage{ulem}
\usepackage{amssymb}

\begin{document}
\title{The role of spin-lattice coupling for ultrafast changes of the magnetic order in rare earth metals}
\author{Beatrice Andres}
\affiliation{Freie Universit{\"a}t Berlin, Fachbereich Physik, Arnimallee 14,
	14195 Berlin, Germany}
\author{Sang Eun Lee}
\altaffiliation[Present address: ]{Department of Physical Chemistry, Fritz-Haber-Institut der Max-Planck-Gesellschaft, Faradayweg 4-6, 14195 Berlin, Germany}
\affiliation{Freie Universit{\"a}t Berlin, Fachbereich Physik, Arnimallee 14, 14195 Berlin, Germany}
\author{Martin Weinelt}
\email[Corresponding author:] {weinelt@physik.fu-berlin.de}
\affiliation{Freie Universit{\"a}t Berlin, Fachbereich Physik, Arnimallee 14,
14195 Berlin, Germany}
\date{\today}

\begin{abstract}
By comparing femtosecond laser-pulse-induced spin dynamics in the surface state of the rare earth metals Gd and Tb we show that the spin polarization of the valence states in both materials decays with significantly different time constants of 15\,ps and 400\,fs, respectively. The distinct spin polarization dynamics in Gd and Tb are opposed by similar exchange splitting dynamics in the two materials. The different time scales observed in our experiment can be attributed to weak and strong $4f$ spin to lattice coupling in Gd and Tb suggesting an intimate coupling of spin polarization and 4f magnetic moment.  While in Gd the lattice mainly acts as a heat sink, it contributes significantly to ultrafast demagnetization of Tb. This helps explain why all optical switching is observed in FeGd- but rarely in FeTb-based compounds.

\end{abstract}

 % control !
% insert suggested keywords - APS authors don't need to do thising.
%\keywords{}

\maketitle

The rare earth metals gadolinium and terbium are important components for materials that show single-shot all optical switching (AOS) \cite{Kirilyuk2010,Kimel2019}.
For this purpose they are combined with transition metals either in ferrimagnetic alloys \cite{Stanciu2007} or as multilayers in artificial ferrimagnets \cite{Lalieu2017}. 
In fact, alloys and multilayers of a wide variety of magnetic materials could be switched but only by applying multiple and circularly polarized laser pulses \cite{Mangin2014,Hadri2016}. In contrast, it turned out highly robust to toggle switch FeCoGd alloys with a single, linearly polarized pulse \cite{Ostler2012}. The only prerequisites are the right range of composition and the concomitantly adjusted magnetic compensation temperature of the alloy \cite{Beens2019}. 
In contrast to the wealth of publications on Gd, there are only few studies reporting AOS in Tb-based samples \cite{Liu2015,Aviles2020}. The absence of AOS in TbFe alloys has been attributed to the strong magnetic anisotropy \cite{Khorsand2013}.

In transition-metal/rare-earth compounds angular momentum exchange between $3d$ and $5d$ spins of the itinerant valence electrons occurs via \textit{inter-}atomic exchange and optically excited (superdiffusive) spin currents, while $5d$ and $4f$ spins of the rare earth couple via \textit{intra-}atomic exchange.  Various theories have been developed to model AOS, for example in Refs.~\onlinecite{Mentink2012,Wienholdt2013,Beens2019,Davies2020}, which all involve the exchange of angular momentum between the antiferromagnetically coupled sublattices  (layers) via exchange scattering, spin-polarized currents or magnons. Here $5d$ and $4f$ spins of the rare earth are considered as a single spin system that couples to the $3d$ spins of the transition metal. However, for Gd disparate dynamics of $5d$ exchange splitting and $4f$ magnetic moments was demonstrated, which challenges strong coupling between optically excited $5d$ spins and not primarily excited $4f$ spins \cite{Frietsch2015,Andres2015}. 
 
This raises the question, how $5d$ and $4f$ spins of rare earth metals couple. Building on recent results \cite{Frietsch2020}, we compare the spin dynamics in pure Gd and Tb metal grown as epitaxial thin films on W(110). Specifically, we follow our study on Gd \cite{Andres2015} and investigate the response of the Tamm-like $d_{z^2}$ surface state of Tb on optical excitation. While the spin polarization of the Gd surface state decreases with a slow time constant $\tau$ of 15\,ps (Ref. \onlinecite{Andres2015}), the spin polarization shows an ultrafast drop with $\tau = 0.4$\,ps on Tb. Since these time constants are almost identical to those of the decay of the $4f$ spin order \cite{Frietsch2020}, the difference between both rare earth metals can be attributed to very weak and very strong $4f$ spin to lattice coupling in Gd and Tb, respectively. Furthermore, our results demonstrate that the spin polarization of the valence states and the $4f$ magnetic moment are intimately linked and show the same ultrafast dynamics. In  contrast, the exchange splitting of the valence states can show disparate dynamics. This suggests that changing the spin polarization of the rare earth valence states is the route to  $4f$ spin  reversal. In Tb this process is hampered by strong $4f$ spin-lattice coupling and efficient ultrafast transfer of angular momentum into the lattice sink.

\begin{figure*}[t]
	\includegraphics[width=0.99\textwidth]{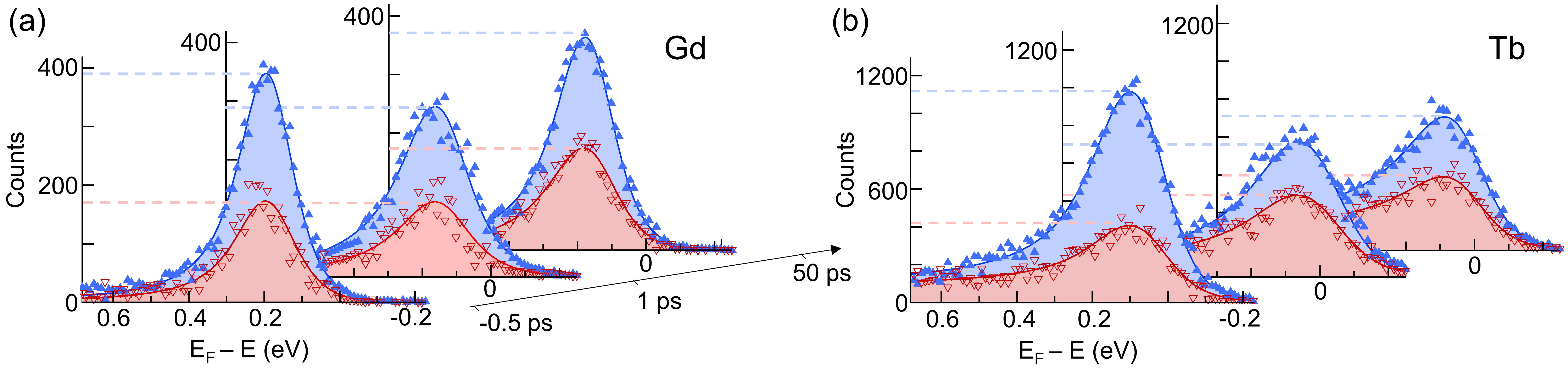}
	\caption{\label{fig:TbGdSpectra} Spin-resolved spectra of the surface sate on Gd (reproduced from \cite{Andres2015}) and Tb before the pump pulse (-0.5\,ps), shortly after pumping (1\,ps), and late after pumping (50\,ps). In the case of Gd, the pump pulse mainly changes the overall peak intensity of both, majority- (blue) and minority-spin peak (red) -- besides slight changes in linewidth and binding energy. In Tb, there is a clear reduction of the spin polarization from the first to the following two spectra, which is easily visible in the ratio between the area covered by the majority- and minority-spin peak.}
\end{figure*}

The experiments were performed with the exchange-scattering spin-detector described in Ref.~\onlinecite{Winkelmann2008} in combination with a cylindrical sector analyzer (CSA 300, Focus) with an angular resolution of $\pm 2.5$$^\circ$. The spin detector uses an oxygen-passivated 6\,ML-thick Fe/W(001) film as scattering target at a backscattering angle of 15$^\circ$. To achieve spin resolution, the magnetization of the Fe target is reversed. This allows us to measure the spin polarization in Tb without the need of a strong magnetic field to reverse its magnetization at liquid nitrogen temperatures.

The Tb and Gd samples (for Gd see also Ref.~\onlinecite{Andres2015}) were grown \textit{in situ} on a W(110) substrate, that was cleaned beforehand by oxygen treatment \cite{Zakeri2010}. Both lanthanides (purity 99.99) were deposited at a rate of 5\,\AA\ per minute from a tungsten crucible in a home-built evaporator at a base pressure of $6 \times 10^{-10}$\,mbar. During evaporation, the substrate was kept at room temperature. Subsequent annealing of the 10\,nm-thick rare earth films was done for 1\,minute at 780\,K for Gd and 850\,K for Tb. While cooling down, Tb was magnetized by field pulses of 20 mT through an air-wound coil. Gd was magnetized at liquid nitrogen temperatures. The magnetization direction was in plane along the [1100] direction. 

As pump pulses, we used the infrared (IR) fundamental (1.58\,eV) of a 300-kHz Ti:sapphire regenerative amplifier (RegA, Coherent). For probe, the ultraviolet (UV) fourth harmonic (6.3\,eV) was created in one frequency-doubling and two sum-frequency-generation steps. IR pump pulses were $s$-polarized to suppress multi-photon photoemission, probe pulses were $p$-polarized. Beams were collinear and the angle of incidence was $45^{\circ}$ off normal along the Gd/Tb[1000] direction. The experiments were performed in two campaigns. For Gd we achieved a temporal resolution (cross correlation of IR pump and UV probe pulses) of 70\,fs while for Tb 160\,fs, as derived from the increase of the hot electron temperature. The IR pump pulses had the same pulse duration of 46 fs.

The majority-spin surface state on Gd and Tb is shown in the spin-resolved spectra of Figure~\ref{fig:TbGdSpectra}. Apart from a $\sim 100$\,meV lower binding energy on Tb, the surface states are identical on both materials. The lower binding energy results in a cutting of the peak by the Fermi function, giving rise to a steeper decreasing flank on the high-energy side of the peak. With the majority-spin state close below $E_F$, the exchange-split minority-spin partner is unoccupied lying above $E_F$ and thus barely measurable in photoemission \cite{footnote_1}.

The majority-spin part shows a spin polarization of 70\% (Gd) and 50\% (Tb) at $90$\,K and thus has a remaining minority-spin population. The difference in spin polarization is ascribed to the different Curie temperatures of 293\,K and 220\,K for Gd and Tb, respectively \cite{Liu2021}.
The temporal behavior of the spin polarization after the pump pulse excites the sample is visualized by the spectra evolving towards the back. The front spectrum was measured at negative pump-probe delay (-0.5\,ps) and is thus unchanged by the pump pulse. The spectra to rear have been measured shortly after the pump at 1\,ps and later at 50\,ps. Comparing the three spectra for Gd in Fig.~\ref{fig:TbGdSpectra}(a), the spin polarization, \textit{i.e.}, the ratio between the majority- and minority-spin intensities stays at a high value. This is indicated by the dashed horizontal lines, which represent the maximum intensity for each peak. Contrasting this, for Tb (Fig.~\ref{fig:TbGdSpectra}(b)) the spin polarization decreases rapidly after pumping and is still low at 50\,ps delay.

\begin{figure}
	\includegraphics[width=0.47\textwidth]{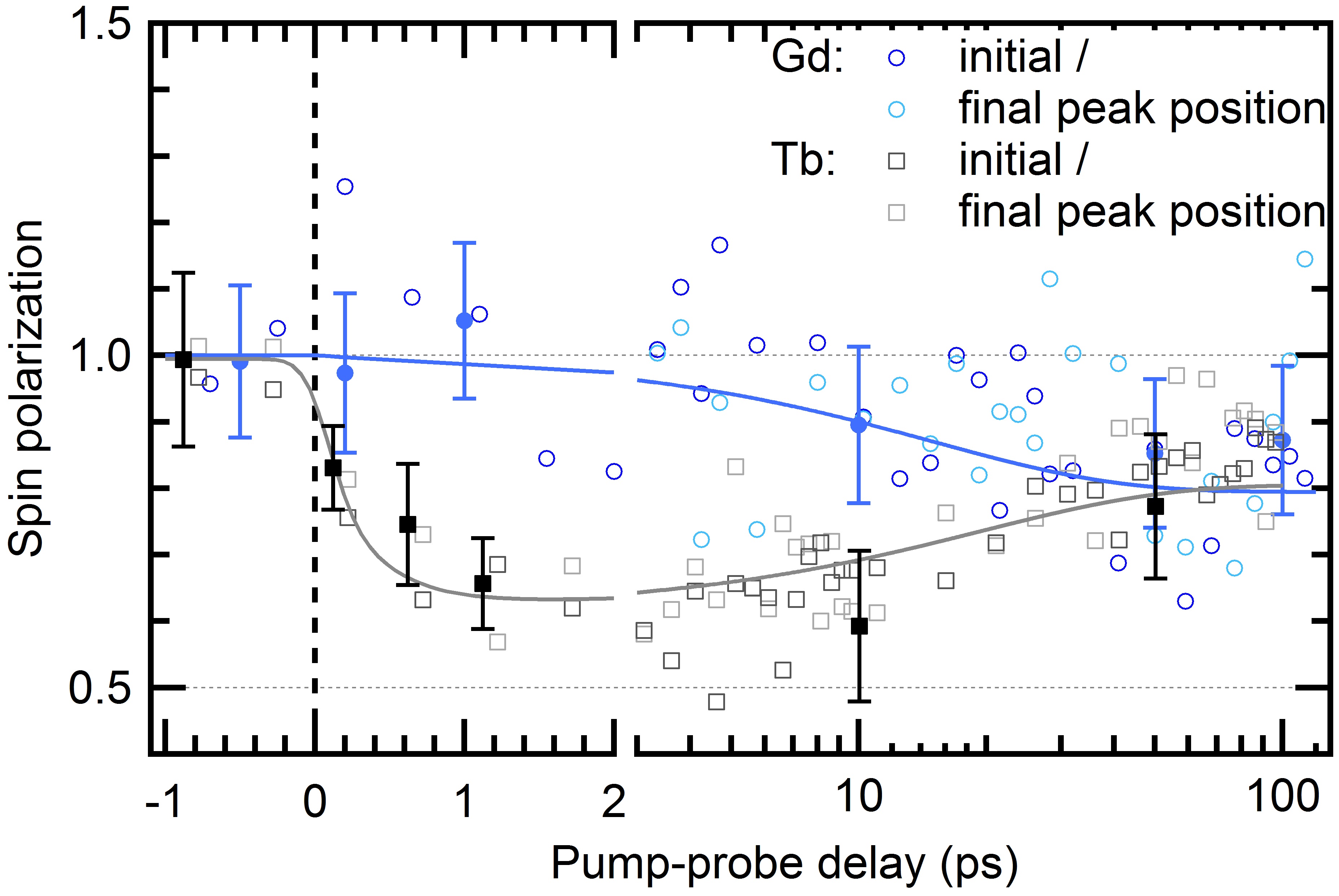}
	\caption{\label{fig:spinPol} Temporal evolution of the normalized spin polarization in the Gd and Tb majority-spin surface state. The spin polarization integrated over the energy-range of the full peak (filled symbols) is complemented by data (open symbols) taken at the initial peak position (-200/-110 meV for Gd/Tb) and final peak position (-120/-30 meV for Gd/Tb) during the pumping process. These energies are also indicated for the Tb case by the dotted and dashed horizontal lines in Fig.\ref{fig:TbMatrix}. While in Gd the spin polarization decreases to only 80\% of the initial value at large pump-probe delays ($\tau_{Gd}=15\pm8$\,ps), Tb shows a much faster response decreasing to 65\% with a time constant $\tau_{Tb}=0.4\pm0.1$\,ps.}
\end{figure}

The complete time evolution of the spin polarization $P$ is presented in Figure~\ref{fig:spinPol}. We calculate $P$ as 
\begin{eqnarray}
P=\frac{I^{\uparrow}-I^{\downarrow}}{I^{\uparrow}+I^{\downarrow}}
\label{Eq:spinpol}
\end{eqnarray}
with $I^{\uparrow / \downarrow}$ being either the full area covered by the majority-/minority-spin peak in an energy scan or the corresponding intensity measured at a certain kinetic energy in a delay scan.
To follow the spin evolution more detailed, we performed delay scans with the CSA set to both the binding energy, the surface state had initially -- before pumping --, and the energy, to which the peak finally shifts during the demagnetization process. These scans are represented by the open symbols in Fig.~\ref{fig:spinPol} (circles for Gd, squares for Tb). The data is completed by the spin polarization evaluated from the energy scans of the full peak (filled circles and squares in Fig.~\ref{fig:spinPol}), some of which were shown in Fig.~\ref{fig:TbGdSpectra} (see Supplementary Material, SM).

In Figure~\ref{fig:spinPol}, it is clearly visible that the spin polarization of the Tb surface state decreases quickly after pumping contrasting the spin dynamics in the Gd surface state. In Gd, we find a very slow decrease on a timescale at which the spin polarization in Tb already recovers. 
We fitted the temporal behavior of the spin polarization using the following function:
\begin{eqnarray}
P_{Tb}(t) & = & P_0 \cdot (1\,-\,A\cdot \int S(t)dt \cdot  (1 - e^{-t / \tau_{\rm{Tb}}}) \cdot e^{-t / \tau_{\rm{r}}}),\nonumber \\
& & 
\label{eq:spinPolfitTb}
\end{eqnarray}
in which $P_0$ is the initial spin polarization, $A$ is the amplitude of demagnetization and $\tau_{\rm{Tb}} (\tau_{\rm{r}})$ are the demagnetization (remagnetization) time constants. The excitation by the laser pulse $S(t)$ was approximated by multiplication with a step function at t = 0 ps. Since we didn't observe the start of a remagnetization within the measured time range in Gd, $P_{Gd}$ was fitted without the exponential remagnetization term.
\begin{eqnarray}
P_{Gd}(t) = P_0 \cdot (1 - \int S(t) dt \cdot A \cdot (1 - e^{-t / \tau_{\rm{Gd}}})).
\label{eq:spinPolfitGd}
\end{eqnarray}
Equations~\ref{eq:spinPolfitTb} and \ref{eq:spinPolfitGd} have been convolved with a Gaussian resembling the time resolution of our laser pulses.

A fit (solid lines) through the delay scans results in very different time constants of $\tau_{\rm{Gd}}=15\pm8$\,ps and $\tau_{\rm{Tb}}=0.4\pm0.3$\,ps. Additionally, the magnitude of the reduction differs by almost a factor of two (reduction by 20\% in Gd, but by 35\% in Tb). The demagnetization appears to be much more efficient in Tb \cite{Frietsch2020,Eschenlohr2014}. Note that the fluence of the pump pulse was chosen to be much lower in Tb (1.2\,mJ/cm$^2$ absorbed pump fluence in the Tb/W(110) sample) to achieve a demagnetization comparable to the one measured in Gd at an absorbed fluence of 3.5\,mJ/cm$^2$. The stronger spin-lattice coupling obviously enhances the demagnetization in Tb, which would result in a much stronger demagnetization in Tb  compared to Gd upon using the same fluence for both materials.

\begin{figure}
	\includegraphics[width=0.47\textwidth]{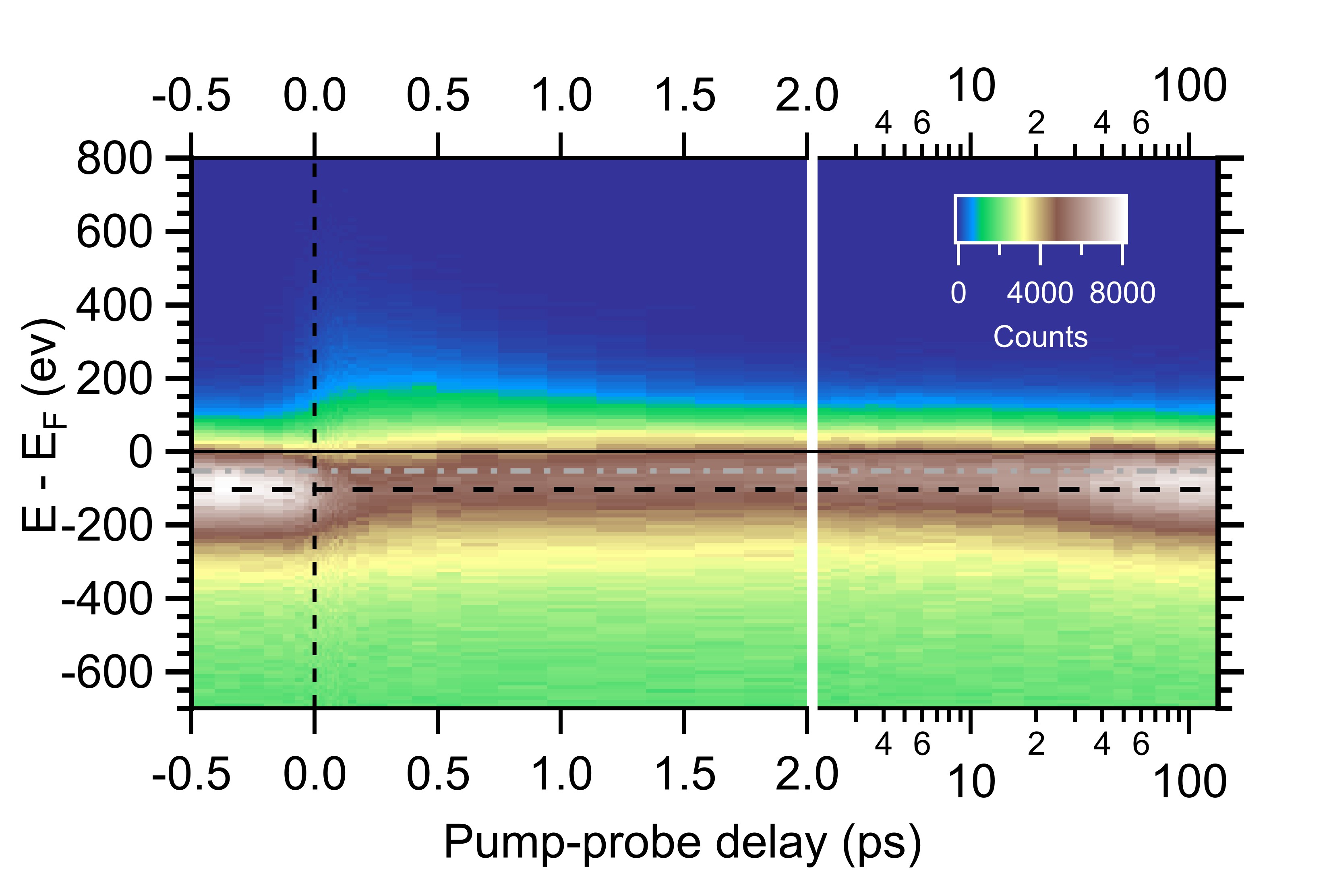}
	\caption{\label{fig:TbMatrix} Spin-integrated spectra of the Tb surface state evolving with increasing pump-probe delay (false-color plot). The surface state is immediately depopulated by the increasing electron temperature, which leads to a broadening of the Fermi function. This also creates an increased intensity above $E_F$. Additionally, the surface state peak shifts towards $E_F$ and recovers to the initial spectrum at $\approx 100$\,ps. The initial and final binding energy reached during the shift is visualized by the black dashed and gray dashed-dotted horizontal lines, respectively.}
\end{figure}

Besides the spin polarization, the demagnetization also affects the exchange splitting and thus the binding energy of the surface state. To determine the change in binding energy, we used the spin-integrated mode of our spin detector by simply flipping the scattering target out of the electron beam path \cite{Winkelmann2008}. The detection is then directly done without back scattering focusing the photoelectrons onto a second channeltron positioned after the entrance lens of the spin detector in forward direction. The spin-integrated mode allows us to measure a delay-series of energy spectra in a comparatively short time without the loss in count rate that is induced by the back scattering used for spin resolution. Such a series for Tb is depicted as false color plot in Fig.~\ref{fig:TbMatrix}. A corresponding graph for the Gd surface state has been published in Ref.~\onlinecite{Andres2015}.

\begin{figure}[t]
	\includegraphics[width=0.45\textwidth]{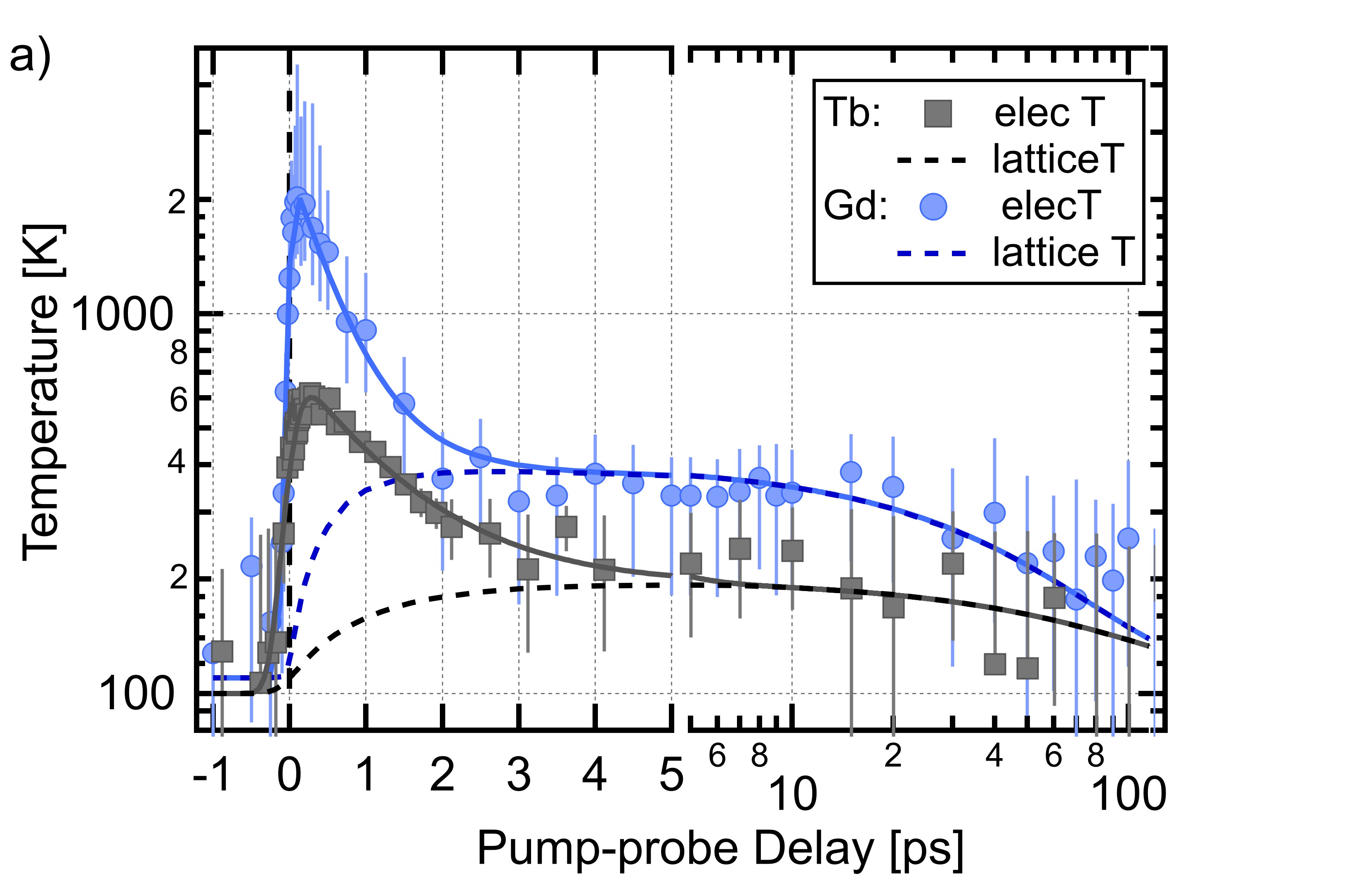}
	\includegraphics[width=0.45\textwidth]{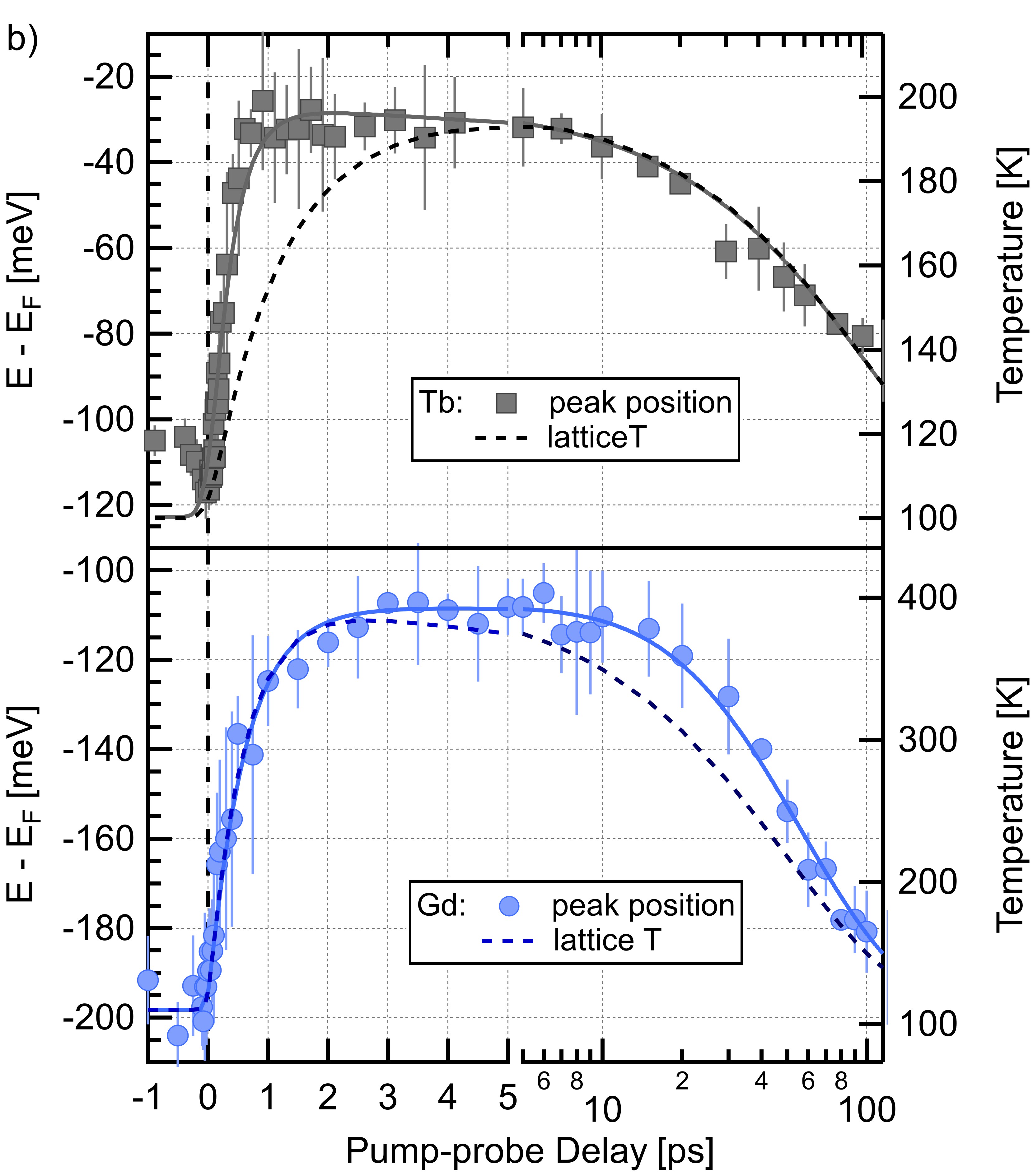}
	\caption{\label{fig:BindEn} Temporal evolution of the electronic temperature and surface state binding energy in Gd and Tb. (a) The electronic temperature $T_e$, represented by the width of the Fermi function in our spectra, increases instantly when the system is pumped at $t=0$. The subsequent drop of $T_e$ can be described by an equilibration with the lattice temperature $T_l$ (dashed lines) in a two-temperature model. (b) The binding energy of the majority-spin surface state decreases during the demagnetization process, which can be attributed to a lowering of the surface state's exchange splitting. The energy shift is on a similar timescale as the change in $T_l$, obtained from (a), represented by the dashed lines. The additional shift, deviating from $T_l$ within the first 5\,ps for Tb and at delays $>3$\,ps for Gd, are caused by the decreasing spin polarization.}
\end{figure}

Figure~\ref{fig:TbMatrix} gives an overview over the whole pumping process. The primarily excited electrons thermalize rapidly  within the temporal evolution of the experiment ($\sim 70$\,fs for Gd), giving rise to a broadening of the Fermi function\cite{Bovensiepen2007}. The broadening in turn increases the photoemission intensity above $E_{\rm{F}}$ and reduces the intensities below $E_{\rm{F}}$. This affects the height of the surface state peak in our spectrum (cf. Fig.~\ref{fig:TbGdSpectra}), which is overlayed by a simultaneous broadening of the peak's linewidth, equally induced by the increasing temperature in the electron system. The surface state peak furthermore shifts towards $E_{\rm{F}}$ (long-dashed horizontal lines in Fig.~\ref{fig:TbMatrix}). Regarding the typical measures for magnetism, the upward binding-energy shift of the majority-spin component of the surface state equals a reduction of the surface state's exchange splitting unless the minority-spin component would likewise shifts upwards. From time-resolved photoemissionon on Gd \cite{Lisowski2005} or temperature-dependent static  scanning tunneling spectroscopy on Tb and Gd \cite{Bode1999} we infer that the unoccupied minority-spin component stays at constant binding energy or shifts towards $E_{\rm{F}}$.
The data in Fig.~\ref{fig:TbMatrix} have been analyzed to deduce the temperature of the hot electron gas and the peak position of the surface state (see Supplementary Material of Ref.~\cite{Andres2015}).  The electronic temperature $T_{\rm{e}}$ is derived from the width of the Fermi function. Figure~\ref{fig:BindEn}a compares  $T_{\rm{e}}$ for Gd (blue) and Tb (gray) as a function of pump-probe delay. Note the linear and logarithmic scales of the abscissa up to and beyond 5\,ps, respectively. The electronic temperature is raised directly within our time resolution of 70\,fs (for Gd) and 160\,fs (for Tb) through the decay of excited hot electrons via inelastic electron-electron scattering. It decreases  through dissipation of heat to the lattice leading to an equilibration of electron and lattice temperature above the $\sim 100$\,K starting temperature within about 4\,ps and 6\,ps at 380 and 220\,K for Gd and Tb, respectively. The different IR pump fluences (3.5 versus 1.2\,mJ/cm$^2$) are reflected in the maximum and equilibrated temperatures, which are higher for Gd than for Tb. We approximate this equilibration by adding the increasing lattice temperature $T_{\rm{l}}$ to $T_{\rm{e}}$. This yields the following fitting functions:
\begin{eqnarray}
T_{\rm{e}}(t) = \int S(t) dt \cdot \left(A_{\rm{el}} \ e^{-t/\tau_{\rm{el}}} + T_{\rm{l}}(t)\right)\\
T_{\rm{l}}(t) = A_{\rm{le}} \left(1 - e^{-t/\tau_{\rm{el}}}\right)e^{-t/\tau_{\rm{d}}},
\label{Eq:2tm}
\end{eqnarray}
where the amplitudes $A_{\rm{el}} >  A_{\rm{le}}$ account for the different heat capacities of electrons and lattice $C_{\rm{e}} < C_{\rm{l}}$. We can fit Eq.~\ref{Eq:2tm} to the temperature shown in Fig.~\ref{fig:BindEn}a after convolving with a Gaussian to account for the time resolution given by the cross correlation of the laser pulses. The sigmoid function resulting from the convolution of step $S(t)$ and Gaussian perfectly resembles the time integral of the Gaussian shaped laser pulse. An additional exponential decay term with time constant $\tau_{\rm{d}}$ was included to account for the cooling of the electrons and lattice through thermal diffusion at large pump-probe delays \cite{footnote_2}. This is legitimate, since thermal diffusion is about two orders of magnitude slower than the equilibration between electron and lattice ($\tau_{\rm{d}} \gg \tau_{\rm{el}}$). The simple relations in Eq.~\ref{Eq:2tm} provide a good modeling of the measured temperatures as demonstrated by the solid lines in Fig.~\ref{fig:BindEn}a. We find a slightly faster decay of $T_{\rm{e}}$ for Gd ($\tau_{\rm{el}} \sim 0.8$\,ps) compared to Tb ($\tau_{\rm{el}} \sim 1.25$\,ps), as expected from the difference between electron and lattice temperatures according to the two-temperature model \cite{Anisimov1974}.

The temporal evolution of the binding energy of the surface state is shown in Fig.~\ref{fig:BindEn}b and can be described by Eq.~\ref{Eq:2tm} as well. The surface state’s binding energy shifts closer to the Fermi level. Fitting the binding energy and  temperature simultaneously, we find that this energy shift occurs in Gd with the same time constant of 0.8\,ps as the lattice temperature increases. It is well described by Eq.~\ref{Eq:2tm} except for a small but significant discrepancy ($\leq 20$\,meV) at larger delays $\geq 5$\,ps when we observe the decrease of the spin polarization of the surface state. For Tb we can in contrast reproduce the binding energy for delays larger than 5\,ps, while we observe a significantly steeper decrease of the binding energy with $\tau = 0.4 \pm 0.2$\,ps, which mimics the decay of the spin polarization (cf. Fig.~\ref{fig:spinPol}) and is three times faster than the increase of the lattice temperature. As discussed below, these deviations are the imprint of the dynamics of the spin polarization on the binding energy of the surface state and reflect the difference in $4f$ spin-lattice coupling between Gd and Tb.

We find three characteristics in the temporal evolution of the surface state after ultrafast optical excitation.

(i) The first is a very fast increase of the electronic temperature. The observed response time is actually set by the 70\,fs (160\,fs) time resolution of the Gd (Tb) experiment. 
The broadening of the Fermi function with increasing $T_{\rm{e}}$ goes hand in hand with a depopulation of the occupied surface state and thus a transfer of majority spins between surface and bulk states. This explains the ultrafast drop (within the laser pulse duration) of the magnetic signal in surface-sensitive second harmonic generation at both the Gd and Tb(0001) surfaces \cite{Lisowski2005,Bovensiepen2007,Melnikov2008}.

(ii) The second observation is the shift of both surface state’s binding energy to lower values, i.e. a reduction of the exchange splitting. On Gd it mainly follows the lattice temperature with a time constant $\tau$ of $0.8 \pm 0.2$\,ps. We observed a comparable response time also in the decay of the exchange splitting of the Gd bulk bands \cite{Carley2012}. On Tb the surface state binding energy responds significantly faster with $\tau = 0.4 \pm 0.2$\,ps. This time constant is again compatible with the decay of the exchange splitting of the Tb bulk bands \cite{Teichmann2015}.

(iii) With the here presented spin-resolved measurements, the difference between Gd and Tb is easily explained by the third feature, the decay of the surface-state spin-polarization. Please note the difference between population of a state and its spin polarization $P$ (see Eq.~\ref{Eq:spinpol}). While the depopulation of the Gd and Tb surface state occurs in accordance with observation (i) during optical excitation within the pump pulse duration, the spin polarization drops very slowly for Gd with $\tau \sim 15$\,ps but decays rapidly for Tb with $\tau = 0.4 \pm 0.2$\,ps (Fig.~\ref{Eq:spinpol}). Obviously, in Tb, the ultrafast decrease of the spin polarization of the surface state (Fig.~\ref{fig:spinPol}) leads to an (within error bars) instantaneous shift of the surface-state binding-energy. This explains the deviation of the binding-energy shift and lattice temperature within the first 3\,ps after laser excitation in Tb (Fig.~\ref{fig:BindEn}b). Accordingly the difference in Gd at delays larger than 3\,ps (Fig.~\ref{fig:BindEn}c) is attributed to the change in spin polarization starting at around 3\,ps (see Fig.~\ref{fig:spinPol}). 

We conclude that the three characteristics observed in this work correspond to three contributions to demagnetization:
(i) The ultrafast optical excitation initiates likewise ultrafast spin transport effects. In our measurements, this manifests as a depopulation of the strongly spin-polarized surface state of Gd and Tb.
(ii) The dissipation of heat between electrons and the lattice gives rise to spin-flip effects mediated by Elliot-Yafet scattering. This is one contribution to the decreasing exchange splitting observed in form of a peak shift in Gd and Tb in this work.
(iii) The very different spin-polarization dynamics, which we observe for Gd and Tb, must be the result of spin-lattice coupling. In Ref.~\onlinecite{Frietsch2020} Frietsch \textit{et al.} have shown that the $4f$ spin systems of Gd and Tb respond likewise on distinct pico- and femtosecond timescales very similar to those observed here for the surface-state spin-polarization. In line with Wietstruk \textit{et al.} \cite{Wietstruk2011} this was attributed to the different $4f$ spin to lattice coupling in both metals. While it is negligible in Gd ($4f^7, L=0$), it is strong in Tb due to its finite orbital momentum $L=3$ of the $4f^8$ configuration. In fact for the latter rare earth, phonon and magnon branches hybridize \cite{Jensen1971}. In line we observed very recently that inelastic electron decay leads to generation of magnon polarons in Tb while phonon and magnon excitations remain decoupled in Gd \cite{Liu2021}. Thus spin polarization and $4f$ magnetization are intimately connected in the rare earth metals \cite{Rex1999,Sandratskii2014} and spin mixing or, synonymously, band mirroring of surface state and most likely all valence bands occurs via magnon emission. 

Returning to all-optical switching we may ask how the $3d$ spins of transition metals and the $5d$ and $4f$ spins of the rare earth metals couple in ferrimagnetic alloys and  artificial ferrimagnets like FeCoGd and Co|Gd multilayers \cite{Stanciu2007,Lalieu2017}.
The comparable response of $4f$ magnetization and $5d$ spin polarization suggests that whenever we modify the spin polarization of the valence electrons, we alter the 
$4f$ magnetic moment. The former is achieved both by spin flips via local exchange scattering of 3d and 5d electrons \cite{Lalieu2017}, \textit{e.g.}, at the interface of $3d$ and $4f$ metals, or by injecting spin currents from the Co layer (or FeCo rich phase) into the Gd layer (or Gd rich phase) \cite{Beens2019,Graves2013}. This likewise explains more efficient demagnetization of the antiferromagnetic (spin spiral) phase of Dy as compared to its low-temperature ferromagnetic phase \cite{Thielemann2017}. 

Why is it easier to switch the magnetization in Gd as compared to Tb compounds? On the ultrafast timescale lattice heating does little affect the $4f$ spin dynamics in Gd but strongly in Tb. Since the lattice is a sink for angular momentum \cite{Dornes2019}, ultrafast spin reversal and angular momentum dissipation compete in Tb but not in Gd. However, switching via spin currents or exchange coupling across interfaces helps to reduce lattice heating and may allow for all optical switching of other artificial ferrimagnets, as recently demonstrated for Co/Pt multilayers \cite{Gorchon2017,Igarashi2020} and Co/Tb stacks \cite{Aviles2020}. An alternative route may be to reduce dissipation channels, \textit{e.g.}, in ferrimagnetic Heusler alloys \cite{Banerjee2020}. 

In summary, the presented data are a further cornerstone to close the puzzle of magnetization dynamics in the rare earth metals. We observe a hierarchy: The 4f magnetization and the spin polarization of the valence bands react conjointly and affect the exchange splitting. Contrary we can change the exchange splitting in Gd by ultrafast optical excitation without affecting the spin polarization of the valence bands and the $4f$ magnetic moment. Thus different contributions of demagnetization are reflected in the exchange splitting (Figs.~\ref{fig:BindEn}b and c). On the one hand, Elliot-Yafet scattering changes the exchange splitting on the timescale of electron-lattice temperature equilibration. On the other hand, the demagnetization of the spin polarization induces a further reduction of exchange splitting, giving rise to the deviations from the lattice temperature dynamics on very different timescales for Gd and Tb.\\

\noindent
SUPPLEMENTARY MATERIAL\\
\noindent In the Supplementary Material, we show the different spin-resolved scans, which have been performed to obtain the data, for the Tb samples. We relate spin-resolved energy-spectra and delay scans to the temporal evolution of the spin-integrated surface band structure and thereby explain the deduction of the spin-polarization curve in Fig.~\ref{fig:spinPol} of the main text.\\

\noindent
AUTHOR'S CONTRIBUTION\\
\noindent
BA and MW designed the experiment. SEL and BA performed the experiment. BA and MW wrote the paper with input from SEL.\\

\noindent
ACKNOWLEDGEMENT\\
\noindent
We acknowledge funding by the Deutsche Forschungsgemeinschaft through CRC/TRR 227 \textit{Ultrafast spin Dynamics}, project A01. The authors like to thank \textit{Focus GmbH} for their steady support of the CSA analyzer over two decades.\\

\noindent
AUTHOR'S DECLARATION\\
\noindent
The authors have no conflicts to disclose.\\

\noindent
DATA AVAILABILITY\\
\noindent
The data are in the Supplementary Material and available on request from the authors. \\

%%\bibliography{GdTb_Dynamics}

\begin{thebibliography}{41}
\bibitem [{foo({\natexlab{a}})}]{footnote_1}%
	\BibitemOpen
	\href@noop {} {} ({\natexlab{a}}),\ \bibinfo {note} {it is generally possible
		to populate the unoccupied minority-spin component of the surface state with
		the pump pulse and ionize it with a second laser pulse in two-photon
		photoemission. Nevertheless, such a two-photon photoemission measurement
		yields count rates at least a factor of thousand lower compared to direct
		photoemission. Thus, we can assume that our data analysis of the occupied
		majority-spin surface state measured in direct photoemission is not
		compromised by any two-photon photoemission contribution of the unoccupied
		minority-spin surface state component.}\BibitemShut {Stop}%
\bibitem [{foo({\natexlab{b}})}]{footnote_2}%
	\BibitemOpen
	\href@noop {} {} ({\natexlab{b}}),\ \bibinfo {note} {we assume that
		electronic diffusion processes are happening within the timescale of our
		laser pulses and are thus not distinguishable from the excitation process in
		our experiment.}\BibitemShut {Stop}%
\end{thebibliography}


\end{document}